\documentclass[%
reprint, superscriptaddress,amsmath,amssymb,aps,
prl,prstper,floatfix,showpacs]{revtex4-1}

\usepackage[FIGTOPCAP,tight]{subfigure}
\usepackage{amssymb}
\usepackage{amsmath}
\usepackage{graphicx}
\usepackage{array}
\usepackage{epstopdf}
\usepackage[usenames, dvipsnames]{color}

\newcommand{\pare}[1]{\left( #1 \right)}

\newcommand{\cor}[1]{\left[ #1 \right]}

\newcommand{\ave}[1]{\left\langle #1 \right\rangle}

\begin{document}

\title{Noise-Enabled Optical Ratchets}

\author{Roberto de J. Le\'{o}n-Montiel}
\email{roberto.leon@nucleares.unam.mx}
\address{Instituto de Ciencias Nucleares, Universidad Nacional Aut\'{o}noma de M\'{e}xico, Apartado Postal 70-543, 04510 Cd. Mx., M\'{e}xico}

\author{Pedro A. Quinto-Su}
\email{pedro.quinto@nucleares.unam.mx}
\address{Instituto de Ciencias Nucleares, Universidad Nacional Aut\'{o}noma de M\'{e}xico, Apartado Postal 70-543, 04510 Cd. Mx., M\'{e}xico}



\begin{abstract}
In this work we demonstrate single microparticle transport enabled by noise in a one dimensional optical lattice with periodic symmetric potentials and a small constant external force. The one dimensional lattice is implemented by six focused beams with holographic optical tweezers, where a microparticle is trapped in three dimensions. Transport initiates when dynamical disorder is added to the diffracted laser power at each trap ($\pm 30\%$) at a fixed frequency (0 to 35 Hz), while the direction of motion is set by the constant external force. We find that transport is only achieved within a narrow noise frequency range, which is consistent with simulations, and the predicted behavior and observations of noise-induced energy transport in quantum and classical systems.
To our knowledge this is the first direct observation of noise-assisted transport in a colloidal system. 
\end{abstract}

\maketitle

Noise has long been considered as detrimental for energy transport in complex systems. However, recently it has been shown that for certain coherently evolving systems, noise can indeed enhance their transport efficiency \cite{mohseni2008,plenio2008,rebentrost2009,caruso2009,kassal2012,roberto2014}. This fascinating phenomenon, coined environment-assisted quantum transport, has been experimentally observed in systems where controllable noise has been introduced in order to enhance the transfer efficiency of electronic \cite{leon2015} and optical \cite{biggerstaff2015,viciani2015,caruso2016} signals. Interestingly, recent theoretical studies have suggested that this effect may be observed in purely classical systems \cite{roberto2013} and, more importantly, that it could be exploited to enhance the transport of microscopic objects across potential energy landscapes \cite{spiechowicz2014,spiechowicz2015,spiechowicz2016}.

Transport across an array of potentials has been achieved in the micro- and nanoscale domain by means of ratchet systems, where movement of a particle is mediated by a combination of a periodic external force and asymmetric potentials which privilege motion in one direction while hindering in the opposite \cite{faucheux1995,lee2005PRL,arzola2011,hasegawa2012,huidobro2013,gomers2006,gomers2008,salger2009}. These asymmetric potentials represent the ratchet and the pawl in the classical Smoluchowski-Feynman ratchet \cite{smoluchowski1912,feynman1963,hanggi1996}, while the periodic force would represent the Brownian perturbations. There is also the possibility of symmetric potentials synchronized with an external force, which breaks the temporal (rather than spatial) symmetry of the system producing directed motion \cite{reimann2001,flach2000,zheng2001}.

In this contribution, we report on the transport of a single microparticle across a one-dimensional symmetric optical lattice by means of dynamical disorder or noise. The optical potentials are created by focused beams (optical tweezers) that trap the microparticle in three dimensions. Movement of the particle is then enabled by introducing random fluctuations in the power of each individual trap, changing the depth of the potentials at a fixed frequency, $f$, which can take different values between $0$ and $35$ Hz. Finally, in order to guarantee a directed motion of the particle, a weak external force is included in the system. This force is smaller than the one necessary to make the particle escape the potentials and hence it is not sufficient to create transport by itself. The system resembles a tilted Smoluchowski-Feynman ratchet \cite{reimann2002}, where a constant external force is added to the potentials, slightly tilting them in the direction of the force.

{\it Model--} We consider a Brownian particle moving in a dynamically-disordered one-dimensional potential landscape (Fig. 1A), comprising $N$ closely-spaced Gaussian potential wells (that mimic the optical potentials), of the form \cite{lee2005,pelton2004}
\begin{equation}
V\pare{x}=-V_{0}\pare{t}\sum_{n=0}^{N}\exp\cor{-\frac{\pare{x-nL}^{2}}{2\sigma^2}},
\label{Eq:potential}
\end{equation}
where $V_{0}\pare{t}$ and $\sigma$ stand for the depth and width of the wells, and $L$ is the separation between them. Because in the experiment noise is introduced by random changes in the power at each trap (trap depth), fluctuations in the depth of the wells may be described by
$V_{0}\pare{t} = V_{0}\cor{1 + \phi\pare{t}}$,
with $V_{0}$ being the average depth of the wells and $\phi\pare{t}$ a Gaussian random variable with zero average, i.e. $\ave{\phi\pare{t}}=0$, where $\ave{\cdots}$ denotes stochastic averaging.

The motion of the Brownian particle in a potential such as the one in Eq. (\ref{Eq:potential}) can be well described by the Langevin equation, in the overdamped limit \cite{volpe2012}, as
\begin{equation}
\dot{x} = -\frac{1}{\gamma}V^{'}_{\text{eff}}\pare{x} + \sqrt{2k_{B}T\gamma}\xi\pare{t}.
\label{Eq:Langevin}
\end{equation}
Here, $\gamma$ characterizes the friction of the particle which is immersed in liquid, and $\sqrt{2k_{B}T\gamma}\xi\pare{t}$ the thermal noise due to random collisions with the surrounding fluid molecules. $\xi\pare{t}$ stands for a Gaussian Markov process with zero average, $k_{B}$ is the Boltzmann constant, and $T$ the temperature of the system. Notice that, in Eq. (\ref{Eq:Langevin}), we have defined an effective potential $V_{\text{eff}}\pare{x} = V\pare{x} - x\delta F$, where $\delta F$ is a weak, constant external force. Because of its construction, one may identify Eq. (\ref{Eq:Langevin}) as a tilted Smoluchowsky-Feynman ratchet \cite{reimann2002}, with the important difference that, in our system, the perturbation $\delta F$ is used only to guarantee a directed motion of the particle and not to create transport by itself.

\begin{figure}[t!]
   \begin{center}
   \includegraphics[width=8.25cm]{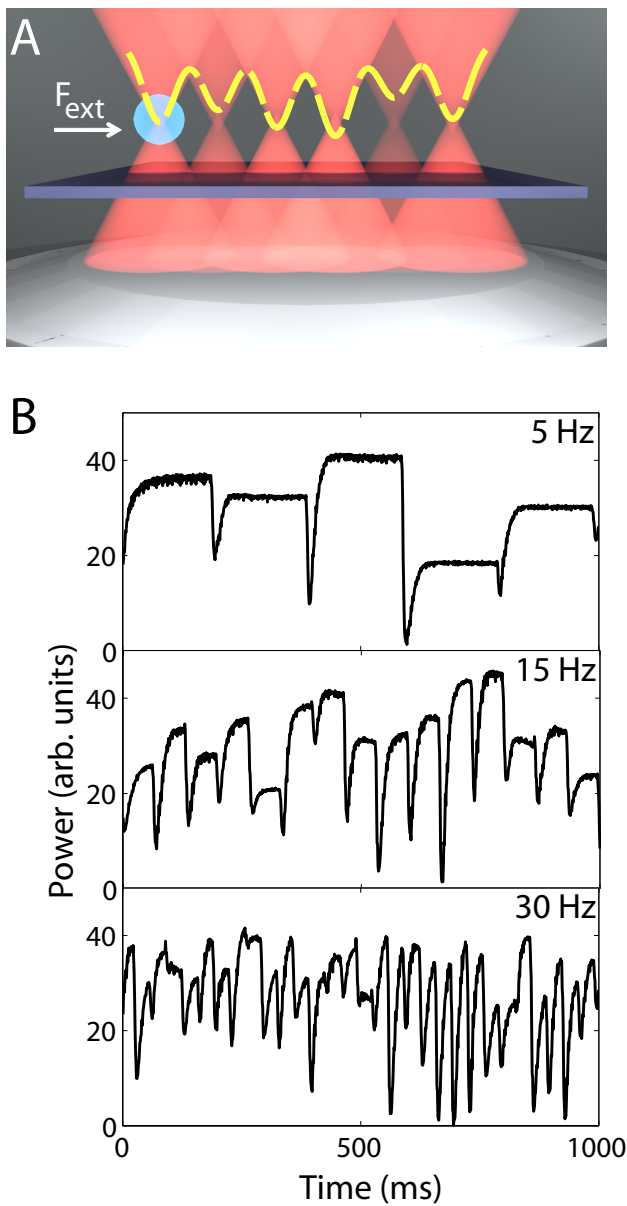}
   \end{center}
   \caption{(A) Schematic representation of the optical potentials. (B) Measured power at one trap at 1000 fps for noise at $f=$ 5, 20 and 30 Hz.}
\end{figure}

{\it Experiment--} The ratchet system is implemented with standard holographic optical tweezers setup where the trapping laser has a wavelength of 1064 nm. The optical potentials are created by adding a two-dimensional phase to the trapping beam with a spatial light modulator (SLM, Holoeye Pluto NIR-2). 
The structured laser beam is simultaneously focused at six spots with a spatial separation of $1.88\pm 0.05~\mu$m by a 100x microscope objective (NA 1.25) in a custom inverted microscope with a piezo-electric stage. 
The total diffracted power into all of the potentials is 26 mW (transmitted by the microscope objective). Noise is introduced to the potentials with random changes in the optical power at a frequency between 0 and 35 Hz with a standard deviation of $\pm 30\%$.
The colloid sample is water with silica microbeads (Bangs Laboratories) that have a mean diameter $2R$ of 2.47$~\mu$m contained within two microscope coverslips (No. 1, 0.13-0.16 mm thick) separated by $\sim 100~\mu$m.

The digital holograms for the six periodic optical potentials are calculated with the Gershberg Saxton algorithm (GS) \cite{gs} and the weighted GS algorithm \cite{wgs}, where the amplitudes are varied randomly with different standard deviations between 5 and $40\%$ while keeping the mean power constant. The intensity for each resulting trap array is measured and the holograms are classified by the measured uniformity in steps of $5\%$. In this way we create videos by arranging the calculated holograms randomly at each uniformity at frame rates between 0.5 to 35 Hz.

The videos are proyected into the SLM while a high speed camera (Photron SA 1.1) records the dynamics of the back reflected beam at 1000 fps. 
Figure 1B shows the integrated power at one trap driven at different frequencies. We observe that each time the frame changes there is flickering, which results in a lowering of diffracted power to the trap. The measured characteristic rise/fall times (defined at 10-90$\%$ of the transition) are: $23.9 \pm 5.3$ ms and $4.9 \pm 1.2$ ms respectively. The time where power is below $10\%$ is  $4.6 \pm 2.2$ ms, in addition we found that that the power dropped to less than $1/4$ of the mean for $36\%$ of the transitions. 

The time to reach a stationary value of the intensity is about 30 ms, that corresponds to a frequency of 33 Hz which is close to an upper bound for the noise frequency. As the frequencies increase the flickering time becomes comparable to the period of the cycle, so that while the intensity is rising the pattern switches before it can reach the stationary value, effectively lowering the mean power delivered to the trap. This also has the effect of rising the standard deviation of the time series for the optical power at each trap. 
Hence, videos at the highest frame rates of 35 Hz with the highest uniformity in the holograms yield the lower boundary for the standard deviation which is about $\pm 30\%$. In this way, at every video frame rate (or noise frequency) we choose the hologram set that yields a variation of $\pm 30~\%$ while adjusting the diffracted power to keep the mean power in the time series constant at all the frequencies.

The constant external force is introduced by dragging the piezo-electric stage at a constant rate of $10~\mu$m/s. Hence the force is the drag experienced by the spherical particle which is given by the Stokes expression $F_d = 6 \pi \eta R v$, where $\eta$ is the liquid viscosity and $v$ is the drag speed.  The drag has to be corrected as the particle is near the bottom boundary at a height $h$ of 16$~\mu$m with the Faxen correction, which to the third order of $r=R/h$ yields $F=F_d/(1-9r/16 +r^3 /8)$ \cite{leach2009}, and has a contribution of $+4\%$ compared to the Stokes expression.

The force necessary to leave the potentials is given by the escape velocity. The measured escape velocity for different patterns with the highest uniformity is $52\pm 12~\mu$m/s. Therefore the external force of $10~\mu$m/s is not enough to induce transport with a static pattern, even for the potential wells that have the lowest power. 
The effect of flicker, where for some events the intensity drops to very low values, the particle can be essentially free for about 5 ms that would result in a displacement of 50 nm (drag speed of 10$~\mu$m/s) which is not enough to reach a neighboring potential well.

\begin{figure}[t!]
   \begin{center}
   \includegraphics[width=8cm]{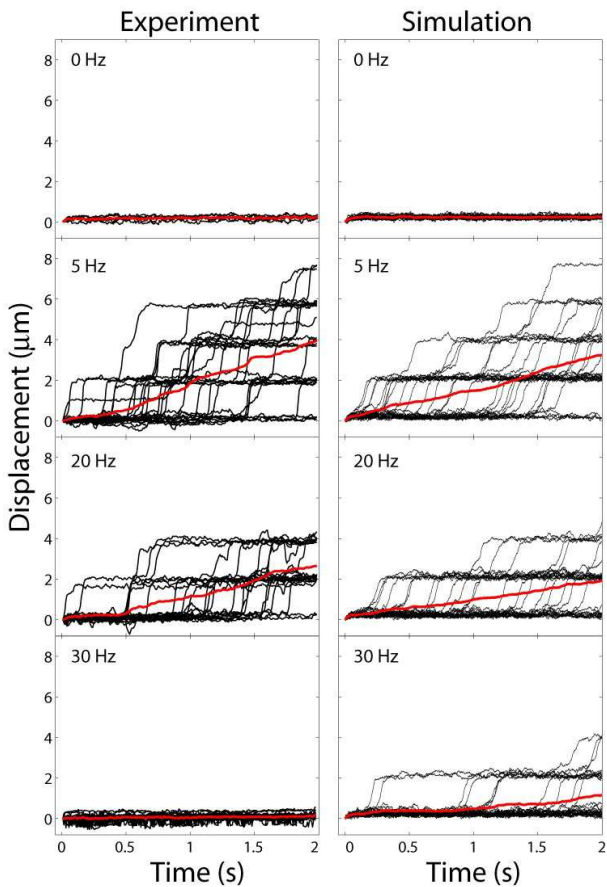}
   \end{center}
   \caption{Left column: Measured microparticle trajectories for noise at $f=$ 0, 5, 20 and 30 Hz. Right column: Simulated trajectories at $f=$ 0, 5, 20 and 30 Hz. The red line represents the average trajectory over 30 different realizations. }
\label{Fig:eff}
\end{figure}

The experiments are done with noise frequencies $f$ between 0 and 35 Hz, and the dynamics of the microparticle are recorded at 125 fps during 60 seconds for each $f$ value. 
A sample of the measured trajectories for selected frequencies are shown in the left column of Fig. 2. The right column shows the simulated trajectories for the same noise frequencies using the model described by Eq. (\ref{Eq:Langevin}). The thick red line represents the average over all trajectories, and the slope of that line is the mean transport speed.  
We observe that, as expected, there is no transport for a static pattern since the drag speed is smaller than the escape velocity.
As the noise frequency increases the optimum transport speed is reached at 5 Hz for the experiment and simulations, while for larger noise frequencies the transport diminishes and stops at 30 Hz in the experiment, whereas the simulation still shows transport with very low transport speeds.

\begin{figure}[t!]
   \begin{center}
   \includegraphics[width=8.25cm]{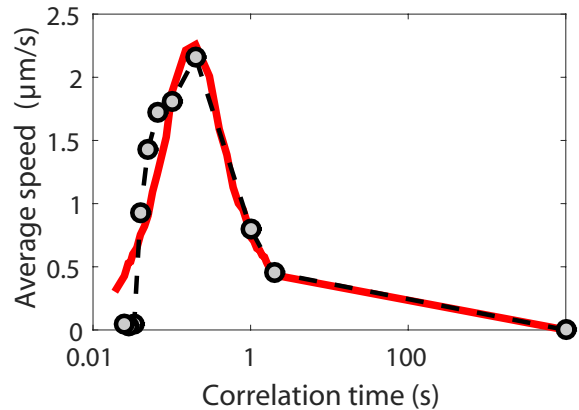}
   \end{center}
   \caption{Average speed of the particle as a function of the noise correlation time $\tau = 1/f$. Experimental results: round symbols (dotted line to guide the eye), were obtained by averaging the speed of the particle over 30 realizations. The red solid line represents the simulated average speed of the particle numerically obtained for the same number of realizations.}
\label{Fig:eff}
\end{figure}

Figure 3 shows the measured average transport speeds (round symbols) as a function of the noise correlation time, which we define as the inverse of the noise frequency $\tau =1/f$. The continuous red line represents the results for the simulations which agree reasonably well in the whole range with a small discrepancy at the highest frequencies. 
The results show that whether noise-assisted transport occurs depends on a competition of time scales. In particular, we can compare the configurational relaxation time of the particle---the time $\tau_{cr} \simeq \eta R^3/(2k_{B}T)$ needed for the particle to diffuse across its own radius \cite{einstein_book,philipse_book}---and the noise correlation time $\tau$. At short correlation times (high frequencies), the optical potentials change so fast, compared to the dynamics of the system, that the particle does not ``feel'' any change in the energy landscape. In this case, the particle is likely to remain trapped in each realization, thus leading to a small average speed of about $0.4\;\mu\text{m}/s$. When increasing the correlation time $\tau$, the average speed of the particle grows up to $2.3\;\mu\text{m}/s$. Interestingly, the maximum value of the particle's average speed is reached when the noise correlation time is of the same order as the configurational relaxation time of the particle, which for our experiment corresponds to $\tau_{cr} \simeq 0.22$ s. This is essentially the same value for $\tau$ at the maximum transport speed. 
In this situation, the particle diffuses at the same rate as the potentials change, which effectively increases the probability of the particle to escape from one potential to the other. Finally, for larger correlation times, the average speed of the particle drops rapidly to zero. This is due to the fact that the correlation time of the noise becomes so large that the optical potentials do not change during each measurement, resulting in a system that is no longer affected by noise.

Noise-assisted transport has been previously explained as the suppression of coherent quantum localization through noise, bringing the detuned quantum levels into resonances and thus facilitating energy transfer \cite{rebentrost2009,kassal2012}. 
The results presented here show that this phenomenon can be observed in a broader class of systems.
This opens interesting routes towards new methods for enhancing the efficiency of the ratchet mechanism, from particle sorting systems to efficient molecular motors. In this way, a phenomenon initially conceived in a quantum scenario has shown to apply as well in classical macroscopic systems, widening the scope of possible quantum-inspired technological applications.

{\it Author contributions--}
RJLM proposed the idea and performed the simulations. PAQS designed and implemented the experiment. Both authors wrote the manuscript.

{\it Acknowledgement--}
Work partially funded by the following grants: DGAPA-UNAM (PAPIIT IN 104415).


\begin{thebibliography}{XX}

\bibitem{mohseni2008} M. Mohseni, P. Rebentrost, S. Lloyd, and A. Aspuru-Guzik, J. Chem. Phys. \textbf{129}, 174106 (2008)..

\bibitem{plenio2008} M. Plenio and S. Huelga, New J. Phys. \textbf{10}, 113019 (2008).

\bibitem{rebentrost2009} P. Rebentrost, M. Mohseni, I. Kassal, S. Lloyd, and A. Aspuru-Guzik, New J. Phys. \textbf{11}, 033003 (2009).

\bibitem{caruso2009} F. Caruso, A. W. Chin, A. Datta, S. F. Huelfa, and M. B. Plenio, J. Chem. Phys. \textbf{131}, 105106 (2009).

\bibitem{kassal2012} I. Kassal and A. Aspuru-Guzik, New J. Phys. \textbf{14}, 053041 (2012).

\bibitem{roberto2014} R. de J. Le\'{o}n-Montiel, I. Kassal, and J. P. Torres, J. Phys. Chem. B \textbf{118}, 10588 (2014).

\bibitem{leon2015} R. de J. Le\'{o}n-Montiel, M. A. Quiroz-Ju\'{a}rez, R. Quintero-Torres, J. L. Dom\'{i}nguez-Ju\'{a}rez, H. M. Moya-Cessa, J. P. Torres and J. L. Arag\'{o}n, Sci. Rep. \textbf{5}, 17339 (2015).

\bibitem{biggerstaff2015} D. N. Biggerstaff, R. Heilmann, A. A. Zecevik, M. Gr\"{a}fe, M. A. Broome, A. Fedrizzi, S. Nolte, A. Szameit, A. G. White, and I. Kassal, Nat. Commun. \textbf{7}, 11282 (2016).

\bibitem{viciani2015} S. Viciani, M. Lima, M. Bellini, and F. Caruso, Phys. Rev. Lett. \textbf{115}, 083601 (2015).

\bibitem{caruso2016} F. Caruso, A. Crespi, A. G. Ciriolo, F. Sciarrino, and R. Osellame, Nat. Commun. \textbf{7}, 11682 (2016).

\bibitem{roberto2013} R. de J. Le\'{o}n-Montiel and J. P. Torres, Phys. Rev. Lett. \textbf{110}, 218101 (2013).




\bibitem{spiechowicz2014} J. Spiechowicz, P. H\"{a}nggi, and J. Luczka, Phys. Rev. E \textbf{90}, 032104 (2014).

\bibitem{spiechowicz2015} J. Spiechowicz and J. Luczka, Phys. Scr. \textbf{165}, 014015 (2015).

\bibitem{spiechowicz2016} J. Spiechowicz, J. Luczka, and P. H\"{a}nggi, Sci. Rep. \textbf{6}, 30948 (2016).




\bibitem{faucheux1995} L. P. Faucheux, L. S. Bourdieu, P. D. Kaplan, and A. J. Libchaber, Phys. Rev. Lett. \textbf{74}, 1504 (1995).

\bibitem{smoluchowski1912} M. von Smoluchowski, Phys. Zeitshur. \textbf{13}, 1069 (1912).

\bibitem{feynman1963} R. P. Feynman, \emph{The Feynman Lectures on Physics, Vol. 1.} (Addison-Wesley, Massachusetts, 1963).

\bibitem{hanggi1996} P. H\"{a}nggi and R. Bartussek, Lecture Notes in Phys. \textbf{476}, 294 (1996).

\bibitem{hasegawa2012} Y. Hasegawa and M. Arita, J. Royal Soc. Interface \textbf{9}, 3554-3563 (2012).


\bibitem{lee2005PRL} S.-H. Lee, K. Ladavac, M. Polin, and D. G. Grier, Phys. Rev. Lett. \textbf{94}, 110601 (2005).

\bibitem{arzola2011} A. V. Arzola, K. Volke-Sep\'{u}lveda, and J. L. Mateos, Phys. Rev. Lett. \textbf{106}, 168104 (2011).

\bibitem{huidobro2013} P. A. Huidobro, S. Ota, X. Yang, X. Yin, F. J. Garc\'{i}a-Vidal, and X. Zhang, Phys. Rev. B \textbf{88}, 201401 (2013).


\bibitem{gomers2006} R. Gommers, S. Denisov, and F. Renzoni, Phys. Rev. Lett. \textbf{96}, 240604 (2006).

\bibitem{gomers2008} R. Gommers, V. Lebedev, M. Brown, and F. Renzoni, Phys. Rev. Lett. \textbf{100}, 040603 (2008).

\bibitem{salger2009} T. Salger, S. Kling, T. Heckling, C. Geckeler, L. Morales-Molina, and M. Weitz, Science \textbf{326}, 1241-1243 (2009).

\bibitem{reimann2001} P. Reimann, Phys. Rev. Lett. \textbf{86}, 4992 (2001).

\bibitem{flach2000} S. Flach, O. Yevtuchenko, and Y. Zolotaryuk, Phys. Rev. Lett. \textbf{84}, 2358 (2000).

\bibitem{zheng2001} Z. Zheng, G. Hu, and B. Hu, Phys. Rev. Lett. \textbf{86}, 2273 (2001).

\bibitem{reimann2002} P. Reimann, Phys. Rep. \textbf{361}, 57-265 (2002).


\bibitem{lee2005} S.-H. Lee and D. Grier, J. Phys.: Condens. Matter \textbf{17}, S3685-S3695 (2005).

\bibitem{pelton2004} M. Pelton, K. Ladavac, and D. G. Grier, Phys. Rev. E \textbf{70}, 031108 (2004).

\bibitem{volpe2012} G. Volpe and G. Volpe, Am. J. Phys. \textbf{81}, 224 (2013).

\bibitem{gs} R. W. Gerchberg and W. O. Saxton, Optik \textbf{35}, 237 (1972).

\bibitem{wgs} R. Di Leonardo, F. Ianni, and G. Ruocco, Opt. Exp. \textbf{15}, 1913-1922 (2007). 

\bibitem{leach2009} J. Leach, H. Mushfique, S. Keen, R. Di Leonardo, G. Ruocco, J. M. Cooper, and M. J. Padgett, Phys. Rev. E \textbf{79}, 026301 (2009).

\bibitem{einstein_book} A. Einstein, \emph{Investigations on the theory of the Brownian movement} (Dover, New York, 1956).

\bibitem{philipse_book} A. P. Philipse, \emph{Notes on Brownian Motion} (Utrecht University, 2011)


\end{thebibliography}
\end{document}